\documentclass[12pt]{article}
\usepackage[totalwidth=450pt, totalheight=590pt, centering]{geometry}

\usepackage{amsmath,amssymb,mathrsfs,revsymb}
\usepackage{bbm} 

\usepackage[dvips]{graphicx}

\usepackage[dvips,bookmarks=true]{hyperref} 
\hypersetup{pdfstartview=FitH,pdfhighlight=/O,colorlinks=false}

\newcommand{\be}{\begin{equation}}
\newcommand{\ee}{\end{equation}}
\newcommand{\ben}{\begin{equation*}}
\newcommand{\een}{\end{equation*}}

\newcommand{\mc}[1]{\mathcal{#1}}

\newcommand{\p}{\partial}

\title{Semiclassical regime of Regge calculus and spin foams}

\author{
Eugenio~Bianchi\footnote{\texttt{e.bianchi@sns.it}} {\it ${}^{ab}$}~~and~~Alejandro~Satz\footnote{\texttt{alejandro.satz@maths.nottingham.ac.uk}} {\it ${}^{c}$}\\[.35em]
\small{\textit{${}^a$Scuola  Normale Superiore,
Piazza dei Cavalieri 7, I-56126 Pisa,  Italy}}\\
\small{\textit{${}^b$CPT Luminy, 
 Universit\'e de la M\'editerran\'ee, F-13288 Marseille, France}}\\
\small{\textit{${}^c$School of Mathematical Sciences, University of Nottingham, Nottingham NG7 2RD, UK}}
}

\date{\small \today}

\begin{document}
\maketitle

\begin{abstract}
Recent attempts to recover the graviton propagator from spin foam models involve the use of a boundary quantum state peaked on a classical geometry. The question arises whether beyond the case of a single simplex this suffices for peaking the interior geometry in a semiclassical configuration. In this paper we explore this issue in the context of quantum Regge calculus with a general triangulation. Via a stationary phase approximation, we show that the boundary state succeeds in peaking the interior in the appropriate configuration, and that boundary correlations can be computed order by order in an asymptotic expansion. Further, we show that if we replace at each simplex the exponential of the Regge action by its cosine -- as expected from the semiclassical limit of spin foam models -- then the contribution from the sign-reversed terms is suppressed in the semiclassical regime and the results match those of conventional Regge calculus.

\begin{flushleft}
PACS: 04.60.Pp; 04.60.Nc
\end{flushleft}

\end{abstract}

\section{Introduction}\label{sec:introduction}

Spin foam models \cite{Rovelli:2004tv,Baez:1997zt,Oriti:2001qu,Perez:2003vx} are a discrete, combinatorial and background-independent framework for developing a quantum theory of gravity as a sum over histories. A main challenge for this approach is to demonstrate that the theory has a well-defined semiclassical regime in which results agree with conventional perturbative theory \cite{Donoghue:1994dn,Burgess:2003jk}. Facing this challenge involves two different kinds of problems: first, the conceptual problems of defining what is understood by semiclassical limit, and how to connect the different languages and conceptual structures of background-independent and perturbative theories establishing some kind of ``dictionary'' between them; second, the computational problem of finding whether explicit calculations of the same semiclassical quantities on both sides of the dictionary really match. 

The last few years have seen great progress on both of these fronts \cite{Ashtekar:2006yw}. The starting point for this progress was the suggestion by Rovelli \cite{Rovelli:2005yj,Bianchi:2006uf} of a procedure for computing the graviton propagator from Loop Quantum Gravity (LQG) \cite{Rovelli:2004tv,Thiemann:2007zz,Ashtekar:2004eh} with the dynamics implemented covariantly in terms of a spin foam model. On the calculational side, the key ingredients are a boundary semiclassical spin network state peaked on large spins and an analytic expression for the large spin asymptotics of the spin foam vertex amplitude \cite{Barrett:1998gs}; on the conceptual side, the framework is the boundary state formalism discussed in \cite{Rovelli:2004tv} and in \cite{Oeckl:2003vu,Oeckl:2005bv,Oeckl:2005bw,Colosi:2007bj,Colosi:2008fv}, which prescribes how to compute observables in the boundary of a spacetime region with a path integral over the interior region only. If $\hat{\mathcal{O}}_1$, $\hat{\mathcal{O}}_2$ are local boundary geometry observables (such as areas, dihedral angles, $3$-volumes or lengths 
\cite{Rovelli:1994ge,Ashtekar:1996eg,Ashtekar:1997fb,Major:1999mc,Thiemann:1996at,Bianchi:2008es}) acting on a space of spin networks $s$ , then the expectation value for their correlation in a boundary geometry $q$ is given in the boundary state formalism by
\begin{equation}\label{SFobs}
 \langle \hat{\mathcal{O}}_1\,\hat{\mathcal{O}}_2\rangle_q=\frac{\sum_s\,\, W[s]\, \hat{\mathcal{O}}_1\,\hat{\mathcal{O}}_2\, \Psi_q[s]}{\sum_s\,\, W[s]\,\Psi_q[s]}\,,
\end{equation}
where $\Psi_q[s]$ is the boundary quantum state, a spin network functional peaked on the classical boundary configuration $q=(h,K)$ (intrinsic and extrinsic geometry), and $W[s]$ is the boundary functional, representing a sum over interior spin foams $F$ bounded by $s$:
\begin{equation}\label{SFW}
 W[s]=\sum_{F,\,\partial F=s}\prod_f A_f(F)\prod_e A_e(F)\prod_v A_v(F)\,.
\end{equation}
Here $A_f$, $A_e$ and $A_v$ are the amplitudes that our spin foam model assigns respectively to the faces, edges and vertices of the spin foam, which depend on the spins $j_f$ and the intertwiners $i_e$ that label faces and edges for a given $F$. The graviton propagator can be extracted from the connected part of (\ref{SFobs}).

This expression has been thoroughly analyzed for the Barrett-Crane \cite{Barrett:1997gw} spin foam model \cite{Bianchi:2006uf,Livine:2006it,Christensen:2007rv,Bianchi:2007vf,Alesci:2007tx,Alesci:2007tg} and has motivated as well the recent introduction of new models \cite{Engle:2007uq,Livine:2007vk,Engle:2007qf,Freidel:2007py,Engle:2007wy}. However, the analysis has been so far restricted to the case where the boundary state has support only on a single graph corresponds to the boundary of a single simplex, with the internal spin foam consisting only in the vertex amplitude that is dual to this simplex. This is an unsatisfactory state of affairs because then the only spin foam variables that are summed over are the boundary spins and intertwiners, which are explicitly peaked by the boundary state on their classical values. Therefore it could be argued that the nonperturbative nature of the theory plays no role in the calculation. A procedure that started with a more general triangulation with many internal vertices, and summed over internal variables in a truly nonperturbative way, with the semiclassicality being enforced only at the boundary variables by the boundary state, would be a much more nontrivial check of the ability of the nonperturbative theory to recover semiclassical results. The question is, then: is semiclassicality at the boundary enough to enforce semiclassicality at the interior for a general triangulation? 

A second  problem, which could be viewed as a particular aspect of the first one, is the ``cosine problem'':  it is known that the large spin limit of the Barrett-Crane vertex amplitude is related to the cosine of the Regge action for the corresponding simplex \cite{Barrett:1998gs}, while the conventional Regge path integral involves only the positive exponential of the action\footnote{This is analogous to what happens in three dimensions for the Ponzano-Regge model \cite{PonzanoRegge:1968}.}. It was argued in \cite{Rovelli:2005yj,Bianchi:2006uf} in the context of a single simplex that the negative exponential does not contribute to the result because it does not match the phase of the boundary state, and its contribution is suppressed. But can something similar occur for the case of a general triangulation, where there are many internal simplices contributing each a similar cosine factor?

These are the questions we set out to investigate in this paper, using Regge calculus \cite{Regge:1961px,Loll:1998aj,Regge:2000wu,Hamber:2007fk} as a testing ground. The Regge equivalent of the spin foam boundary observable correlation (\ref{SFobs}), for a given arbitrary triangulation, is:
\begin{equation}\label{obsRE}
 \langle \hat{\mathcal{O}}_1\,\hat{\mathcal{O}}_2 \rangle_q=\frac{\int \prod_i \mathrm{d}L_i\, W(L_i)\,\hat{\mathcal{O}}_1\,\hat{\mathcal{O}}_2\,\Psi_q(L_i)}{\int \prod_i \mathrm{d}L_i\,W(L_i)\,\Psi_q(L_i)}\;,
\end{equation}
where now the observables $\hat{\mathcal{O}}_1$, $\hat{\mathcal{O}}_2$ are functions of boundary edge lengths $L_i$, $\Psi_q(L_i)$ is a boundary state peaking the boundary variables on a classical simplicial geometry $q$, and $W(L_i)$ is given by the integral over internal edge lengths of a product over simplices.
\begin{equation}
 W(L_i)=\int \prod_{n}\mathrm{d}l_n \prod_{\sigma}\,W_\sigma(l_n^\sigma,L_i^\sigma)\;,
\end{equation}
\begin{equation}\label{eiS}
W_\sigma(l_n^\sigma,L_i^\sigma)=\mu_\sigma(l_n^\sigma,L_i^\sigma)\,\exp \left[ \frac{i}{\hbar} S_\sigma(l_n^\sigma,L_i^\sigma) \right]  \;.
\end{equation}
Here $\sigma$ labels internal simplices, $l_n$ are internal edge lengths and $(l_n^\sigma,L_i^\sigma)$ are those internal and (if any) boundary edges that belong to simplex $\sigma$, $S_\sigma$ is the Regge action for simplex $\sigma$, and $\mu_\sigma(l_n^\sigma,L_i^\sigma)$ is the contribution of simplex $\sigma$ to the integration measure $\mu$ over internal and external edges.

The goals of this paper are twofold. Firstly, we analyze the semiclassical regime of expression (\ref{obsRE}) with a stationary phase asymptotic expansion, and we show that the leading contribution to the integral comes from the configuration where all the internal variables take the correct classical values (i.e. the particular internal configuration that solves the equations of motion for the chosen boundary conditions). The expansion is first done formally as an $\hbar\to 0$ limit, but afterwards its regime of physical validity is identified with a condition on the boundary state. This is a confirmation that the boundary state formalism can enforce semiclassicality for a general triangulation. This is a key hypothesis of the graviton propagator calculations that is difficult to investigate in the spin foam setting. Regge calculus, as a better-defined and understood model for quantum gravity that shares many structural similarities with spin foams, provides an ideal arena for testing and checking the validity of this hypothesis.

Secondly, recall that Regge calculus is actually conjectured to be intimately related to the semiclassical regime of spin foam models, via the ``cosine problem'' mentioned above. Accordingly, we investigate the consequences of replacing (\ref{eiS}) by a similar expression that is conjectured to result from the large spin limit of the spin foam vertex amplitude corresponding to each simplex:
\begin{equation}\label{cosS}
W_\sigma(l_n^\sigma,L_i^\sigma)=\mu_\sigma(l_n^\sigma,L_i^\sigma)\left[P_\sigma(l_n^\sigma,L_i^\sigma) \cos\big(S_\sigma(l_n^\sigma,L_i^\sigma)+\frac{\pi}{4} \big)+D_\sigma(l_n^\sigma,L_i^\sigma) \right] \;,
\end{equation}
where the term $D_\sigma$ is dominant but slowly varying; this ansatz is motivated by the results on the asymptotics of the Barrett-Crane model \cite{Barrett:1998gs,Baez:2002rx,Barrett:2002ur,Freidel:2002mj}. We show that, remarkably, the results of this replacement when the expression is introduced in (\ref{obsRE}) are the same as those obtained from the original Regge expression: the differences between (\ref{eiS}) and (\ref{cosS}) represent deeply quantum configurations that drop out from the semiclassical regime of boundary observables. This is an important step towards showing that the correct semiclassical limit can be indeed recovered from spin foam models, making contact with perturbative quantum gravity a less unmanageable goal.

The paper is organized as follows: Section \ref{sec:2} is dedicated to the purely Regge part of the calculation. Subsection \ref{sec:2.1} introduces the classical Regge action, the path integral for the quantum theory, and the semiclassical boundary state we will consider. Subsection \ref{sec:2.2} contains the explicit evaluation of boundary observables in Regge calculus, showing with a stationary phase analysis that the Regge path integral peaks on a single semiclassical configuration. Subsection \ref{sec:2.3} discusses the particular case of flat space and how to deal with the translational invariance that arises. Then Section \ref{sec:3} repeats the main calculation for the modified path integral resulting from (\ref{cosS}), showing that the semiclassical results are unchanged and therefore that, under certain assumptions, the semiclassical regime of spin foams matches that of Regge calculus. Section \ref{sec:4} is a discussion of the results and suggestions for further work.

Throughout this paper we work in four-dimensional space with Euclidean signature. We use units with $c=1$ but keep $\hbar$ and $\kappa=8\pi G$ explicitly in order to keep track of semiclassical expansions. The Planck length is then $l_P=\sqrt{\hbar \kappa}$.

\section{Semiclassical regime of Regge calculus}\label{sec:2}

\subsection{Regge calculus and boundary state}\label{sec:2.1}
The Regge action for Riemannian four-dimensional simplicial gravity \cite{Regge:1961px,Hartle:1981cf,Friedberg:1984ma,MTW:1973} can be written as
\begin{equation}\label{action}
 S=\frac{1}{\kappa} \sum_a A_a \phi_a \;.
\end{equation}
Here $a$ ranges over the faces of the triangulation, with $A_a$ representing the area and $\phi_a$ the deficit angle at each face\footnote{Most of our results would be unchanged if we included in the action a cosmological term summing over the volume of simplices, $-\frac{\Lambda}{\kappa}\sum_\sigma V_\sigma$. We do not include explicitly this term for the sake of brevity, but we will point out any point where it would make a difference.}. The deficit angles are defined by
\begin{equation}
 \phi_a = 2\pi - \sum_\sigma \theta_a^{\,\sigma} \;,
\end{equation}
where $\theta_a^{\,\sigma}$ is the internal dihedral angle at face $a$ of simplex $\sigma$, and the sum is over all simplices meeting at face $a$. If the manifold has a boundary, the action takes the same form with the deficit angles for boundary faces given instead by
\begin{equation}
 \phi_a = \pi - \sum_\sigma \theta_a^{\,\sigma} \;.
\end{equation}

The action (\ref{action}) is to be considered a function of the edge lengths ${l_i}$, with $i$ ranging over all edges. The Regge equations of motion obtained from the variation of the action are:
\begin{equation}\label{eom}
 \sum_a \frac{\partial A_a}{\partial l_i} \phi_a = 0\;.
\end{equation}

The quantum theory is normally defined, for a given triangulation, by the path integral over all edge lengths in the triangulation\footnote{``Problem 57'' in \cite{Wheeler:1964rgt}. 
} \cite{Loll:1998aj,Regge:2000wu,Hamber:2007fk}:
\begin{equation}\label{Z-def}
 Z=\int \prod_i\mathrm{d}l_i\, \mu(l_i)\,\exp\left(\frac{i}{\hbar}S(l_i)\right) \;,
\end{equation}
where $\mu(l_i)$ is a suitable measure function \cite{Hamber:1997ut,Jevicki:1985ta,Menotti:1995ih,Menotti:1996tm} and some fixed asymptotic  boundary condition is understood. Following the framework outlined in the Introduction, we use instead the boundary formalism in which the lengths integrated over are only those in the boundary ($L_i$) and the interior ($l_n$) of a closed, compact region, with a boundary quantum state $\Psi_q(L_i)$ included in the path integral:
\begin{equation}\label{Z-int}
 Z_q=\int \prod_i \mathrm{d}L_i\, \prod_n\mathrm{d}l_n\, \mu(l_n,L_i)\,\exp\left(\frac{i }{\hbar}S(l_n,L_i)\right)\, \Psi_q(L_i)\;.
\end{equation}
Expectation values of boundary observables $\hat{\mathcal{O}}$ acting on the space of boundary states are defined by
\begin{equation}
 \langle\, \hat{\mathcal{O}} \,\rangle_q  = \frac{1}{Z_q} \int \prod_i \mathrm{d}L_i\, \prod_n\mathrm{d}l_n\, \mu(l_n,L_i)\,\exp\left(\frac{i}{\hbar}S(l_n,L_i)\right)\; \hat{\mathcal{O}}\; \Psi_q(L_i)\;.
\end{equation}

The choice of boundary state can be taken as equivalent to the choice of asymptotic boundary conditions for the full path integral; in fact, as we discuss in Section \ref{sec:4}, the boundary state can be \textit{defined} as the result of integrating out the external edge lengths in (\ref{Z-def})\footnote{A key property making this viable is that the action $S$ and the measure $\mu$ are both local, splitting respectively into a sum and a product over simplices of single-simplex actions and measures. Therefore talking about the ``internal'' action and measure of a region is well-defined.}. We refer to \cite{Dittrich:2007wm} for a discussion of this procedure in the case of linearized Regge calculus on flat space. Aside from the discussion in Section \ref{sec:4}, in this paper we will not be concerned with the first-principles derivation of the boundary state. We take instead the ``phenomenological'' route that assumes the boundary state to be a Gaussian in the fluctuations (plus eventual higher-order corrections) around some specified classical boundary configuration $\{L_m^0\}$. This follows the approach used in \cite{Rovelli:2005yj,Bianchi:2006uf,Christensen:2007rv,Bianchi:2007vf,Alesci:2007tx,Alesci:2007tg} for spin foam models and in \cite{Bianchi:2007vf} for Regge calculus with a single simplex. For a discussion of the role of the boundary state in the three-dimensional case with a single Ponzano-Regge vertex see \cite{Speziale:2005ma,Livine:2006ab,Livine:2007mr,Bonzom:2008xd}.
 
More concretely, the recipe we use for the boundary state is as follows. Given a triangulation\footnote{Throughout this paper by triangulation we mean only the topological triangulation, also known as the skeleton.} and the 3d subset of it that is the boundary, we want the boundary state to be a function of the boundary lengths $\{L_i\}$ that selects a particular semiclassical regime for the quantum theory: one that is peaked on a particular classical solution. For this we first specify a set of values for the boundary lengths, $L_i=L_i^0$. When used as Dirichlet boundary conditions in the equations of motion (\ref{eom}), there may be for the internal lengths either (a) no solution, (b) a unique solution, (c) a discrete set of solutions, or (d) a continuum of solutions. We are not interested in the case (a), because our goal is to write a state that peaks us on a classical solution; let us therefore assume that we choose $\{L_i^0\}$ so there is at least one solution. 

Focusing on case (b) first, call the unique internal solution $l_n=l_n^0(L_i^0)$. Take the derivative of the internal action with respect to the boundary edge lengths, and evaluate it at the classical solution:
\begin{equation}
K_i^0 = \left.\frac{\partial S}{\partial L_i}\right|_{\substack{L_i=L_i^{0}\\l_n=l_n^0}}\quad.
\end{equation}
This is the boundary extrinsic curvature of our chosen configuration. An alternative definition of $K_i^0$ is the derivative of the Hamilton function with respect to the boundary length $L_i$. The Hamilton function $S^H$  is the internal action evaluated at the classical solution, as a function of boundary variables, and the equivalence of both definitions comes from the expression
\begin{equation}
\frac{\partial S^H}{\partial L_i}=\frac{\partial S}{\partial L_i}+\frac{\partial S}{\partial l_n}\frac{\partial l_n}{\partial L_i}\;,
\end{equation}
where the last derivative is evaluated through the dependence of internal variables on external ones once equations of motion are solved. When evaluated at a classical solution the second term vanishes and the equality is established. 

The Gaussian boundary state that we use is then given by:
\begin{equation}\label{Gauss}
 \Psi_q(L_i)= C\, \exp\left[-\frac{1}{2 \hbar \kappa}\,\alpha_{ij}\,(L_i-L_i^0)\, (L_j-L_j^0)\right] \exp\left[-\frac{i}{\hbar}\,K_i^0\,(L_i-L_i^0)\right] \;,
\end{equation}
where the matrix $\alpha_{ij}$ is symmetric and has positive-definite real part, $C$ is a normalization constant, and summation over $i,j$ indexes running over boundary edges is understood. The reason the phase of the Gaussian involves $K_i^0$ is that $K_i$ is the canonically conjugate variable to $ L_i$; the state $\Psi_q$ is then peaked in both ``position'' and ``momentum'', as a good semiclassical state should. See \cite{Rovelli:2005yj,Bianchi:2006uf} for a more extended discussion, and the next subsection of this paper for a discussion of the consequences of using the wrong phase.

If we are in case (c), then we need only a slight modification to the above procedure. The boundary lengths $L_i=L_i^0$ are not enough to specify a solution. To get a state peaked on a particular solution of the many compatible with the Dirichlet condition, we choose the solution we are interested in, calculate its extrinsic boundary curvature $K_i^0$ as above, and construct the Gaussian boundary state (\ref{Gauss}) in the same way. Then this boundary state selects the semiclassical regime peaked around the particular solution chosen.

Case (d), finally, is what happens when one solution compatible with the given boundary lengths includes a patch of flat space: a region where one or more internal vertices of the triangulation can be translated keeping the physical geometry invariant, thus producing for some of the edge lengths a continuum of values representing the same simplicial geometry. In this case, the action evaluated in all these solutions will be the same -- the Hamilton function of the configuration -- and therefore there is a well-defined extrinsic boundary curvature $K_i^0$ given by its derivative with respect to the boundary length $L_i^0$. The corresponding Gaussian boundary state peaks us on the single interior \textit{geometry} compatible with the given boundary conditions, despite not fixing a unique set of values for the internal edge lengths. We will see in Section \ref{sec:2.3} that this is enough for a well-defined evaluation of observables.

To conclude this section, note that if the $\alpha_{ij}$ coefficients are scale-independent (i.e. they are invariant under a global rescaling of lengths) then the Gaussian fluctuations for both conjugate variables vanish asymptotically for $L_0\gg\sqrt{\hbar\kappa}$, where $L_0$ is a typical length scale of the chosen classical solution:
\begin{equation}\label{PsiDisp} 
\frac{\Delta L_i}{L_i^0}\sim\frac{\Delta K_i}{K_i^0}\sim\frac{\sqrt{\hbar\kappa}}{L_0}\longrightarrow 0\,\,,\quad L_0\gg\sqrt{\hbar\kappa}\;.
\end{equation}
Thus our state is sharply peaked on a classical configuration, with small relative dispersion, when the typical length scale of this configuration is much larger than the Planck length. The ratio of the Planck scale to the boundary scale will serve as a physical expansion parameter in our calculation, and we will evaluate only the lowest nontrivial order of observables in this expansion. This justifies us in using a Gaussian boundary state; higher-order corrections can be taken into account order by order in the asymptotic expansion using an improved boundary state, as discussed in \cite{Bianchi:2007vf}.

\subsection{Stationary phase evaluation of observables}\label{sec:2.2}

In this section we show explicitly how the computation of boundary geometry observables proceeds within our framework. To the order of accuracy allowed by restricting the boundary state to the Gaussian form (\ref{Gauss}), a boundary observable is given by:
\begin{align} \label{obs22}
\langle \hat{\mathcal{O}}\rangle_q &= \frac{1}{Z_q}\int \prod_{n,i} \mathrm{d}l_n\,\mathrm{d}L_i\,
 \mu(l_n, L_i)\,\exp\left( \frac{i}{\hbar} S(l_n,L_i)\right)  \hat{\mathcal{O}}\nonumber\\
&\times\exp\left[-\frac{1}{2 \hbar \kappa}\,\alpha_{ij}\,(L_i-L_i^0)\, (L_j-L_j^0)\right] \exp\left[-\frac{i}{\hbar}\,K_i^0\,(L_i-L_i^0)\right] \; .
\end{align}
Here $ \hat{\mathcal{O}}$ may be an intrinsic boundary geometry observable, i.e. a function of boundary lengths $L_i$ that acts simply by multiplication, or an extrinsic boundary geometry observable, i.e. a function of the boundary extrinsic curvature $K_i$ that acts on $\Psi_q$ as a derivative with respect to $L_i$.

Expression (\ref{obs22}) is a well-defined integral, which could in principle be computed numerically (after a prescription for the measure is chosen). We are interested in evaluating it analytically in a semiclassical approximation. The idea is that the boundary state enforces the integral to be peaked in a particular classical internal configuration. A ``picture'' for how this happens can be seen by doing the integral layer by layer (cf.\cite{Barrett:1994ks}): starting by integrating out the boundary lengths that are explicitly peaked on classical values, and then moving towards the interior, with each ``layer'' of edges getting peaked by the immediately previous one in the configuration that solves the inwardly propagating equations of motion. 

Carrying out this procedure explicitly may be feasible for particular examples, but it is complicate and makes difficult the extraction of information about boundary correlations. A much simpler technique that works for general configurations is to use a global stationary phase approximation. This can be done formally as an $\hbar\to 0$ limit, or in other words an asymptotic expansion of the integral for large values of $\hbar^{-1}$. We will show that this peaks the whole integral in the global configuration that solves the equations of motion for the given boundary. An a posteriori examination of the result will exhibit that this formal asymptotic expansion provides in fact a good approximation if a physical condition is satisfied: the ratio of the typical scale selected by  $\Psi_q$  to the Planck scale must be large. The small $\hbar$ stationary phase approximation is a way to shortcut to an asymptotic expansion in this physical parameter. 
 
Examining (\ref{obs22}) we see that if the argument of the exponentials was a purely imaginary quantity we could use a straightforward stationary phase approximation, while purely real (and negative) argument would require a Laplace approximation. But the argument in (\ref{obs22}) is complex, so we need a more careful statement of our approximation. Let us therefore make a brief mathematical detour to state the precise result that we will apply and also the formulas for evaluating successive orders in the asymptotic expansion, beyond the lowest-order expression that can be found in textbooks.

We are interested in estimating an integral of the form
\begin{equation}
F(\lambda)=\int_D \prod_{i=1}^d d x^i\;f(x)\;e^{\displaystyle i \lambda\, Q(x)} 
\label{eq:F}
\end{equation}
for large parameter $\lambda$. In (\ref{eq:F}), $D$ is an open region in $\mathbb{R}^d$, $f$ and $Q$ are two smooth functions $f:D\to \mathbb{R}$ and $Q:D\to \mathbb{C}$. We assume that $f$ vanishes on the boundary of $D$. Let $H_{ij}(x)$ be the Hessian of $Q(x)$:
\begin{equation}
H_{ij}(x)=\frac{\p^2 Q}{\p x^i \p x^j}(x)\;.
\label{eq:Hessian}
\end{equation}
We assume that the imaginary part of the Hessian is positive semi-definite, $\textrm{Im} [H_{ij}(x)]\geq 0$ $\forall x\in D$. As a result of the eventual presence of zero eigenvalues of $\textrm{Im} [H_{ij}(x)]$, we have that the set of minima of $\textrm{Im} [Q(x)]$ is a subset of $D$, which we call $D_0$. Let us assume for the moment that $\mathrm{Re}[Q(x)]$ has at most one stationary point. There are then three possible scenarios: the real part of $Q(x)$ has (i) no stationary point in $D$, or (ii) it has a stationary point $\bar{x}$ in $D$, but it does not fall in the subset $D_0$, or (iii) there is a stationary point $\bar{x}$ falling in $D_0$. 

We can state the following extension of the stationary phase approximation \cite{Erdelyi:1954,vanderCorput:1954}: in the cases (i) and (ii), for large parameter $\lambda$ the function (\ref{eq:F}) decreases faster than any power of $\lambda^{-1}$, $F(\lambda)=o(\lambda^{-N})\;\;\forall N\geq 1$. On the other hand, in case (iii), the function $F(\lambda)$ is of order $\lambda^{-d/2}$ and (assuming $\textrm{Det} [H_{ij}(\bar{x})]\neq 0$) for large $\lambda$ the asymptotic expansion of the integral is\footnote{If there are several isolated stationary points with non-degenerate Hessian within $D_0$, we obtain a sum of contributions of the form (\ref{eq:asymptotics}) from each of them; if there is in addition one or more stationary point outside $D_0$, its contribution is $o(\lambda^{-N})\;\;\forall N\geq 1$ and hence suppressed in comparison with the other ones. The only remaining case of interest, which is a continuum of stationary points within $D_0$, will be considered in the next section.}
\begin{equation}
F(\lambda)=\left(\frac{2\pi}{\lambda}\right)^{\frac{d}{2}} \frac{e^{-i\, \textrm{ind}[H_{ij}](\bar{x})}}{\sqrt{|\textrm{det} H_{ij}(\bar{x})|}}\; e^{\displaystyle i \lambda\, Q(\bar{x})} \;\; \Big(\sum_{n=0}^N a_n \lambda^{-n}+o(\lambda^{-N})\Big)\;.
\label{eq:asymptotics}
\end{equation}
In (\ref{eq:asymptotics}), the index $\textrm{ind}[H_{ij}]$ is defined in terms of the eigenvalues $\mu_j(H)$ as $\textrm{ind}[H_{ij}]=\frac{1}{2}\sum_j\textrm{arg}[\mu_j(H)]$ with $-\frac{\pi}{2}\leq\textrm{arg}[\mu_j(H)]\leq+\frac{\pi}{2}$. The asymptotic expansion coefficients $a_n$ can be expressed in terms of derivatives of the functions $f(x)$ and $Q(x)$ evaluated at the stationary point $\bar{x}$. 

We define $R(x)$ to be the difference between $Q(x)$ and its second-order Taylor polynomial at $\bar{x}$, $R(x)=Q(x)-Q(\bar{x})-\frac{1}{2} H_{ij}(\bar{x})(x-\bar{x})^i(x-\bar{x})^j$. Such function has a Taylor expansion around $\bar{x}$ which starts with a term cubic in $(x-\bar{x})^i$. The coefficient $a_n$ is given by\footnote{The formal expression $f(\bar{x}+\frac{\p}{\p J})$ and the analogous one for $R(x)$ stand for a Taylor expansion of the functions around the point $\bar{x}$ with the fluctuation substituted by a derivative
\begin{equation}
f(\bar{x}+\frac{\p}{\p J})=f(\bar{x})+\sum_{b=1}^{\infty}\frac{1}{b!}f^{(b)}_{i_1\cdot\cdot i_b}(\bar{x}) \frac{\p}{\p J_{i_1}}\cdot\cdot\frac{\p}{\p J_{i_b}}\;.
\label{eq:der}
\end{equation} 
Notice that for fixed $n$ in (\ref{eq:an}) only a finite number of derivatives appear. Here $f^{(b)}_{i_1\cdot\cdot i_b}(\bar{x})$ denotes the $b$-th partial derivative of $f$ with respect to $x_{i_1}\cdots x_{i_b}$, evaluated at $\bar{x}$.}
\begin{equation}
a_n=i^n\sum_{V=0}^{2n}\frac{(-1)^V}{V!(V+n)!} \left.f(\bar{x}+\frac{\p}{\p J}) \big(R(\bar{x}+\frac{\p}{\p J})\big)^V \Big(\frac{1}{2}(H^{-1}(\bar{x}))^{ij} J_i J_j\Big)^{n+V}\right|_{J=0}\;.
\label{eq:an}
\end{equation} 
In particular, the zeroth-order and the first order coefficients are given by
\begin{align}
a_0=&\;f(\bar{x})\;,\\[6pt]
a_1=&\;\;i \,\frac{1}{2} f^{(2)}_{ij}(\bar{x}) \,(H^{-1}(\bar{x}))^{ij} + \nonumber\\
  &+ i\, f^{(1)}_i(\bar{x}) \; R^{(3)}_{jkl}(\bar{x})\, (H^{-1}(\bar{x}))^{ij}\, (H^{-1}(\bar{x}))^{kl}+\nonumber\\
   &+ i\, \frac{5}{2} f(\bar{x})\; R^{(3)}_{ijk}(\bar{x})\, R^{(3)}_{mnl}(\bar{x})\, (H^{-1}(\bar{x}))^{im} \,(H^{-1}(\bar{x}))^{jn}\, (H^{-1}(\bar{x}))^{kl}\;.\nonumber
\end{align}
The coefficients $a_n$ admit a Feynman-diagrammatic representation in terms of propagators $(H^{-1}(\bar{x}))^{ij}$, vertices $R^{(b)}_{i_1\cdot\cdot i_b}(\bar{x})$ with $b\geq3$ and insertions corresponding to derivatives of $f$ at $\bar{x}$. Since only diagrams with $n$ loops contribute to $a_n$, this kind of approximation is called ``loop expansion'' in the context of quantum field theory \cite{Itzykson:1980rh}.

Now we go back to our physical problem to see how these mathematical formulas apply. As we said, $\hbar^{-1}$ plays the role of large parameter $\lambda$ for the stationary phase evaluation of (\ref{obs22}). Our domain $D$ consists in a subregion of $\mathbb{R}^{M+N}$, where $M$ and $N$ are the number of boundary and internal edges in our triangulation; let us call points in this space $x=(L_1,\cdots L_M,l_1,\cdots l_N)$. The subregion is specified by the measure $\mu(L_m,l_n)$, which vanishes for negative values of the lengths, and also for values not satisfying a set of triangular inequalities that make the configuration possible\footnote{We assume that $\mu$ vanishes smoothly at $\p D$ so there are no boundary contributions to the integral.}. The function we need to extremize is
\begin{equation}\label{bigS} Q(l_n,L_i)=S(l_n,L_i)+\frac{i}{2\kappa}\,\alpha_{ij}\left(L_i-L_i^0\right)\left(L_{j}-L_{j}^0\right)- K_i^0\,\left(L_i-L_i^0\right)\;,
\end{equation}
and its Hessian is given in block matrix form by
\begin{equation}\label{matrix}
 \mathbf{H}=\left(
\begin{array}{cc}
 \mathbf{A}+i\mathbf{\alpha} & \mathbf{N}^{\textrm{t}}\\
\mathbf{N} & \mathbf{B}\\
\end{array}\right) 
\end{equation}
where the block separation corresponds to boundary and interior variables, in that order, and
\begin{equation}
 [\mathbf{A}]_{ij}=\frac{\partial^2 S}{\partial L_i \partial L_j}\,,\quad\quad [\mathbf{\alpha}]_{ij}=\frac{\alpha_{ij}}{\kappa}\,, \quad\quad
[\mathbf{N}]_{in}=\frac{\partial^2 S}{\partial L_i \partial l_n}\,,\quad\quad [\mathbf{B}]_{np}=\frac{\partial^2 S}{\partial l_n \partial l_p}\;.
\end{equation}

Our assumption that $\mathrm{Im}[H]$  is positive semi-definite is clearly satisfied thanks to the properties we have stated for $\mathbf{\alpha}$. It is clear that the subset $D_0$ of the domain where $\mathrm{Im}[Q]$ finds its minimum will be given by points of the form
\begin{equation}
 x=(l_n,L_i^0)\;;
\end{equation}
that is, the condition that the imaginary part of $Q$ be minimized imposes the boundary lengths at the extremum to be equal to the classical values specified in the boundary state, and it imposes no further condition on the interior lengths.

Now we look at the real part of $Q(x)$; taking the above condition coming from the imaginary part as given, the stationary points will be the solutions to the equations:
\begin{subequations}
\begin{align}
\frac{\partial S (l_n,L_i^0) }{\partial l_n}& =\;  0 \;\;, \label{extremum2}\\[6pt]
\left.\frac{\partial S (l_n,L_i)}{\partial L_i}\right|_{L_i=L_i^0}-K_i^0 & =\;  0 \;\;.\label{extremum3}
\end{align}
\end{subequations}
The first of these amounts to the equations of motion with boundary conditions ${L_i^0}$. Assuming we are in case (b) defined in the previous subsection, there is a single solution for the internal lengths, $l_n=l_n^0(L_i^0)$. The second equation can then be simultaneously satisfied only if the phase factor of the boundary state, $K_i^0$, matches the extrinsic boundary curvature at the classical solution. A similar thing happens if we are in case (c), the difference being that then the second equation is not just a consistency condition but actively selects the desired classical internal solution out of the several compatible with the boundary lengths. In both cases, if the phase of the boundary state is an arbitrary quantity unrelated to the solution(s) implied by the boundary lengths, we fall into case (i) or (ii) of the theorem and the semiclassical limit is suppressed. This confirms that only a state properly peaked in intrinsic and extrinsic classical geometry can select correctly the semiclassical regime of the theory. We stress that our formalism does not merely peak the path integral in any configuration that solves the classical equations of motion, but only in a specific classical solution, namely the one that solves the equations of motion for the boundary conditions encoded in the boundary state.

We still need to check that $\mathbf{H}(\bar{x})$ is invertible. As far as the Hessian of the Regge action is concerned, it is known \cite{Rocek:1982fr,Rocek:1982tj} that it has in general a zero mode corresponding to global conformal rescalings, and also translational zero modes in the particular case of patches of flat space. The conformal zero mode disappears from $\mathbf{H}$ when we add the $i\alpha$ part coming from the boundary state; physically, this corresponds to the fact that the boundary state fixes the geometry of the boundary and in particular its scale\footnote{The conformal zero mode reflects the fact that, given one classical solution, a family of classical solutions can be found (in absence of matter and of cosmological constant) simply rescaling all the lengths by a given parameter. However, these are \emph{new} solutions: they represent simplicial geometries with observables physically larger or smaller than the original ones. In the quantum theory, the semiclassical regime is identified by a semiclassical boundary state, which is peaked on a \emph{specific} intrinsic boundary geometry with a specific scale; this selects one among the classical solutions related by a rescaling, and therefore the conformal mode disappears from the Hessian when $\alpha$ is included. A different perspective is taken in \cite{Dittrich:2007wm}, where the conformal zero mode is considered a gauge degree of freedom and a gauge-fixing procedure is advocated.}. The translational degeneracy persists, as can be easily checked; this corresponds to the fact mentioned as case (d) in the previous subsection, that if boundary conditions corresponding to an internal solution that is totally or partially flat space are fixed, then there is a continuum of internal classical solutions, which correspond to translations of interior vertices of the triangulation. For the moment we assume that the specific solution selected by our boundary state does not contain flat regions, and then there will be a single extremum with nondegenerate $\mathbf{H}$. The case of flat space is left for the following subsection.

In summary, the point in configuration space from which the integral gets its dominant contributions as specified by the stationary phase approximation is $(L_i^0,\,l_n^0(L_i^0))$: the boundary lengths at their chosen classical values, the internal ones at the classical values determined from the boundary conditions via the equations of motion. 

Let us consider next the evaluation of observables. We focus on intrinsic boundary geometry observables depending only on boundary lengths, but a parallel discussion could be done for extrinsic boundary geometry. Applying our collection of formulas, the expectation value of an observable $\mc{O}(L_i)$ is given by
\begin{equation}
\langle \mc{O}(L_i) \rangle_q = \mc{O}(L_i^0) + \textrm{order $\hbar$ corrections}\;.
\label{obs0}
\end{equation}
This comes from keeping only the zeroth order in the asymptotic expansion both for the numerator and the denominator $Z_q$; notice that dependence on the measure $\mu$ drops out. Note also that, while the ``kinematical'' expectation value of an observable (the result of integrating it against the modulus squared of the Gaussian boundary state over the boundary variables $L_i$) is exactly $\mc{O}(L_i^0)$, the ``dynamical'' expectation value calculated with (\ref{obs22}) acquires corrections of order $\hbar$.

We are specifically interested in \textit{correlations} of boundary observables, which are of order $\hbar$. The simplest example is the connected two-point function for lengths, which is related to the simplicial graviton propagator studied in \cite{Rocek:1982fr,Rocek:1982tj,Hamber:2004ew}. Applying our asymptotic expansions to the first order gives:
\begin{align}
\langle\, L_i\, L_j \,\rangle_q - \langle\, L_i\, \rangle_q \;\langle\, L_j\, \rangle_q  =&\; \hbar \;[\mathbf{H}^{-1}]_{ij} + O(\hbar^2) \\[6pt]
=&\; \hbar \; [(\mathbf{A}+i\mathbf{\alpha}-\mathbf{N}^{\textrm{t}}\, \mathbf{B}^{-1}\mathbf{N})^{-1}]_{ij}+O(\hbar^2)\;.
\label{twopoint}
\end{align}
Recall that $\mathbf{H}$ is a $(M+N)$-dimensional matrix involving second derivatives of the action with respect to boundary variables $L_i$ and internal variables $l_m$. The correlation function involves taking its inverse and selecting from it the $M\times M$ block corresponding to the boundary indices. According to the well-known formulas for inverting block matrices of the form (\ref{matrix}), we obtain the second line of (\ref{twopoint}).

Expression (\ref{twopoint}) has actually a very simple and beautiful meaning. Consider the Hamilton function $S^H(L_i)=S(l_m(L_i),L_i)$: the action evaluated at the classical solution for given boundary lengths $\{L_i\}$, considered as a function only of these boundary lengths via the equations of motion. We have that the second derivatives of the Hamilton function can be written in terms of derivatives of the action evaluated at the classical solution in the following way
\begin{equation}
\frac{\p^2 S^H}{\p L_i \p L_j}(L_i^0) =  \left.\frac{\p^2 S}{\p L_i \p L_j}-\Big(\frac{\p^2 S}{\p l_m \p l_n}\Big)^{-1}\frac{\p^2 S}{\p L_i \p l_m} \frac{\p^2 S}{\p L_j \p l_n}\right|_{l_m=l_m(L_i^0)}\;.
\label{hamilton}
\end{equation}
This identity  can be easily derived using the implicit function theorem, considering the first derivatives of $S(l_m,L_i)$ being set to zero as describing a constraint surface that makes the internal variables $l_i$ an implicit function of the boundary variables $L_i$. Comparing (\ref{hamilton}) with (\ref{twopoint}), we see that the two-point correlation function to lowest order is given by the inverse of an $M\times M$ matrix which is the sum of two terms: the Hessian matrix of the Hamilton function and the correlation matrix of the boundary state
\begin{equation}
\langle\, L_i\, L_j \,\rangle_q - \langle\, L_i\, \rangle_q \,\langle\, L_j\, \rangle_q  = \hbar \, \Big(\frac{\p^2 S^H}{\p L_i \p L_j}(L_i^0)+\frac{i}{\kappa} \alpha_{ij}\Big)^{-1}\;.
\label{2point-H}
\end{equation}
Note that since the action includes a prefactor $\kappa^{-1}$ both matrices scale in the same way: the two-point function is given by $\hbar\kappa={l_P}^2$, multiplied by a dimensionless matrix. For fixed boundary state, the ``scaling with distance'' of the two-point function (the fact that correlations decrease with distance in a way compatible with the $1/r^2$ behaviour of continuum gravity) is hidden in the way (\ref{2point-H}) changes when considering different pairs $(i,j)$ of edges: edges that are further away correlate more weakly than nearby edges. Recovering the exact scaling known from conventional perturbative theory on a regular lattice \cite{Rocek:1982fr,Rocek:1982tj} can be imposed as a condition on the boundary state matrix $\alpha$. This is discussed more extensively in Section \ref{sec:4}.

From the correlation function for lengths (\ref{2point-H}), all two-point correlation functions of local intrinsic boundary geometry observables can be obtained accurately to the lowest order. In particular the area-area correlations, which were computed for the case of a single regular simplex in \cite{Bianchi:2007vf}, can be easily obtained for a general configuration. As is generally the case in semiclassical perturbation theory, the correlation functions are to the lowest order independent of the nonperturbative measure $\mu$. In a Feynman-diagrammatic language, at \emph{tree level} the measure does not contribute, but the \emph{one-loop} $O(\hbar^2)$ correction does depend on the measure. Finding the general form of the second-order result would also involve using an improved boundary state of the kind discussed in \cite{Bianchi:2007vf}.

To conclude this section, let us discuss which is the physical regime in which the first terms of our formal asymptotic expansion provide a good approximation for the observables. For this we compare the computed expectation value of a boundary observable to its classical value, and require that
\begin{equation}
\xi:=\frac{\langle \, \mc{O}(L)\,\rangle_q-\mc{O}(L^0)}{\mc{O}(L^0)}\ll 1\;.
\label{eq:}
\end{equation}
Computing this with the stationary phase asymptotic expansion for a general observable leads to a condition on the classical solution we have chosen to peak upon. Consider a classical solution which oscillates over a length scale $\lambdabar$. From the general form of the expansion coefficients $a_n$ it can be seen that 
\begin{equation}\label{lambdascale}
 \xi\sim\hbar \kappa/\lambdabar^2\;.
\end{equation}
This is analogous to the conditions of validity of the WKB approximation in quantum mechanics \cite{Landau:1960}. Recall in turn that the classical solution is uniquely determined (up to the flat space degeneracy discussed in the next section)\footnote{In case (c), where more than one classical solution correspond to given boundary lengths, the argument that follows keeps on holding since $K_i^0$, which determines the solution in conjunction with $L_i^0$, scales with $L_0$ as well.} by boundary lengths $L_i^0$. In absence of matter and cosmological constant $\Lambda$ (which introduce intrinsic length scales in the classical theory), the Regge action is a homogeneous function of the edge lengths. Under these conditions, a global rescaling of the edge lengths by a factor $L_0$ leads to an identical rescaling in all the internal lengths at the classical solution determined by the new boundary lengths. Hence our parameter $\lambdabar$ will also scale with $L_0$. We conclude that we can take $\hbar \kappa/L_0^2$ as the physical parameter of our expansion, with $L_0$ being a typical scale of our boundary lengths $L_i^0$. Thus the semiclassical regime is characterized by having boundary lengths very large with respect to Planck length:
\begin{equation}\label{L0scale}
 \xi\sim \frac{{l_P}^2}{{L_0}^2}\,\ll 1\;.
\end{equation}
Note that this ratio is precisely the parameter that quantifies the kinematical relative dispersion in the boundary Gaussian according to (\ref{PsiDisp}). If the action includes a cosmological term or matter sources, then (\ref{L0scale}) is not generally true, while (\ref{lambdascale}) keeps on holding; in this case, careful study of the classical solution is needed to identify a good expansion parameter, which is no longer a single global scale of the boundary state.

\subsection{Flat space}\label{sec:2.3}

Let us now go back to (\ref{obs22}) and consider the case we labelled (d) in Section \ref{sec:2.1}: the classical internal solution implied by our boundary data includes patches of flat space. It is easy to construct such solutions; for example, one can start from any solution defined on a given triangulation, and then modify the triangulation adding an extra vertex inside a simplex, specifying the lengths of the new five edges to be such that the ten new triangles that include these edges all have zero deficit angle (a ``one-five'' Pachner move). See \cite{Dittrich:2008ar} for a recent discussion of such configurations. When the classical solution that our boundary data selects includes flat regions in this way, the edge lengths are redundant variables, in the sense that there is a continuum of values for them -- corresponding to translations of some internal vertices of the triangulation -- that correspond to the same physical geometry. As a result, all solutions for the edge variables within this continuum of values have the same value of the action, which is the extremum for the given boundary data. 

If the number of internal vertices that can be translated leaving the Hamilton function unchanged is $q$, then the $N$ internal length variables $l_n$ can be divided in one set of $4q$ variables parametrizing the position of the points in the flat four-dimensional interior regions, and a set of $N-4q$ variables that are in a sense ``orthogonal'' to this subspace. For given values of the first $4q$ variables, there is a unique set of values for the remaining $N-4q$ variables that extremizes the action. Thus it is possible to do first the integrals in the $M$ boundary variables and in $N-4q$ of the internal variables using a well-defined stationary phase approximation of the form described in last section: the boundary variables get peaked in the values $L_i^0$, and the $N-4q$ internal variables get peaked in the values that extremize the action for fixed value of the remaining $4q$ variables\footnote{This is akin to the so-called method of collective coordinates in statistical field theory.}. This leaves only the remaining $4q$-dimensional integral over all the possible flat configurations, which share the same physical geometry. 

Using the notation $x_n$ for the $N-4q$ internal variables integrated over, calling the other $4q$ internal variables $y_n$, and calling $x_n^0(y_n)$ the values of the $x_n$ that extremize $Q$ for given values of $y_n$ (and $L_i^0$, though we leave this dependence implicit), the result we obtain after the first set of integrations is done by stationary phase is:
\begin{equation}\label{flatstat}
\langle \hat{\mathcal{O}} \rangle =\frac{1}{Z_q}\int \prod_{n=1}^{4q}\mathrm{d}y_n\, \frac{\exp\left[\frac{i}{\hbar}Q(L_i^0,x_n^0(y_n),y_n) \right]}{\left[\det \mathcal{H}(L_i^0,x_n^0(y_n),y_n)\right]^{1/2} } 
\sum_j \hbar^j a_j(L_i^0,x_n^0(y_n),y_n)\;.
\end{equation}
where $\mathcal{H}$ stands for the Hessian matrix of second derivatives of $Q$ with respect to the variables $\{L_i,x_n\}$. As usual, the coefficients $a_j$ are defined  by (\ref{eq:an}) with the role of $Q$ being played by (\ref{bigS}) and the role of $f$ by $\mu\,\mathcal{O}$. $Z_q$ is identical to the numerator but omitting $\mathcal{O}$. Note that (\ref{flatstat}) allows well-defined computation of the boundary observables to any desired order; the boundary state peaking on a specific \textit{geometry} is enough and there is no need to ``gauge fix'' in addition the position of the internal vertices in flat regions.

To the zeroth order, we have as before that in the numerator $a_0$
equals the measure $\mu(L_i^0,x_n^0(y_n),y_n)$ times the insertion $\mathcal{O}(L_i^0)$. The second factor does not depend on the $y$ variables, and upon taking it out from the integral we are left with a ``volume'' integral over the flat regions that cancels out with $Z_q$. Hence the result of the previous section
\begin{equation}
\langle \mc{O}(L_i) \rangle_q = \mc{O}(L_i^0) + \textrm{order $\hbar$ corrections}\,.
\label{obs0flat}
\end{equation}
is still true when flat regions are taken into account.

What about the two-point correlation function? Taking (\ref{flatstat}) up to $O(\hbar)$ for this observable leads to
\begin{align}
&\langle L_i\, L_j \rangle_q - \langle L_i\rangle_q \,\langle L_j\rangle_q  =\label{2pointflat1}\\
&\; = \hbar\, \frac{\int \prod_{n=1}^{4q}\mathrm{d}y_n\,\left[\det \mathcal{H}(L_i^0,x_n^0(y_n),y_n)\right]^{-1/2}\, \mu(L_i^0,x_n^0(y_n),y_n)\,\mathcal{H}^{-1}_{ij}(L_i^0,x_n^0(y_n),y_n)}{\int \prod_{n=1}^{4q}\mathrm{d}y_n\,\left[\det \mathcal{H}(L_i^0,x_n^0(y_n),y_n)\right]^{-1/2}\,\mu(L_i^0,x_n^0(y_n),y_n)}\,.\quad\nonumber
\end{align}
The argument used in the previous subsection to relate the Hessian of $Q$ to the Hessian of the Hamilton function still applies to $\mathcal{H}$, even though it does not include derivatives with respect to the $y_n$ variables, because these are precisely the variables the action is independent of. Hence in (\ref{2pointflat1}) we can replace $\mathcal{H}^{-1}_{ij}$ by $\left( \frac{\p^2 S^H}{\p L_i \p L_j}(L_i^0)+i\,\alpha_{ij}/\kappa\right) ^{-1}$, which is transparently independent of the $y_n$ variables. We can therefore take it out from the integral in the numerator and then cancel out the remaining ``volume'' integrals, leaving us with
\begin{equation}\label{hamilton-flat}
\langle\, L_i\, L_j \,\rangle_q - \langle\, L_i\, \rangle_q \,\langle\, L_j\, \rangle_q  = \hbar \, \Big(\frac{\p^2 S^H}{\p L_i \p L_j}(L_i^0)+\frac{i}{\kappa} \alpha_{ij}\Big)^{-1}\;.
\end{equation}
which is the same expression we had in the previous subsection. The conclusion is that the first-order correlation function can be found directly from the Hamilton function of the configuration, when there are flat regions just as when there are not.

An interesting implication of this is that the leading order of the correlation function in a flat configuration is \textit{triangulation-independent}. All our calculations so far have assumed a given fixed triangulation. But consider now that we choose a different triangulation of the interior region, one for which the internal solution selected by the fixed boundary state is (at the level of physical geometry) the same one as before; the differences between them amount only to a series of Pachner moves performed strictly within flat regions of the original solution. Since this transformation leaves the Hamilton function unchanged\footnote{This is because the Hamilton function depends only on the classical simplicial geometry and not on the triangulation and the assignment of edge lengths \emph{separately}.}, it also leaves unchanged the first-order quantum correlations of boundary geometrical observables. We do not know if this result is preserved at higher orders in the asymptotic expansion; in principle, there seems to be no reason to expect it. 

These results are of great interest for the ongoing discussion on the meaning and scope of triangulation independence in spin foam models. They suggest, for example, that spin foam observables computed with the boundary formalism for a configuration with five spin foam vertices (which amount to five simplices in the original triangulation) might equal those for a single vertex, at least to the first order in a semiclassical expansion peaked on a flat configuration.

\section{Semiclassical regime  of spin foam models}\label{sec:3}

In this section we undertake the extension of our results for Regge calculus to the context of spin foams. A full discussion of the semiclassical limit of spin foam models would require starting with a model and attempting to compute correlations of observables on a semiclassical state, as outlined in equations (\ref{SFobs}) and (\ref{SFW}). We should then show that -- by some reason analogous to the stationary phase analysis of Regge calculus -- the contribution coming from the desired internal semiclassical geometry dominates the state sum, and the results can therefore be related to those of conventional perturbative theory. 

Fulfilling this outline in its entirety is beyond the scope of this paper. We focus instead on examining one particular problem that might hinder its completion -- the ``cosine problem'' mentioned in the Introduction. The conjectured semiclassical limit of the spin foam vertex amplitude is not merely the exponential of the Regge action that appears in the Regge path integral, as would seem to be needed to get a perfect correspondence with simplicial gravity, but instead features the cosine of the Regge action plus a term with no clear geometrical interpretation \cite{Barrett:1998gs,Baez:2002rx,Barrett:2002ur,Freidel:2002mj}. It was shown in \cite{Rovelli:2005yj,Bianchi:2006uf} that this does not thwart the correct semiclassical limit for the case of a single simplex, but doubt remained as to whether this success would survive when more general triangulations are considered. In this section we address this question within the general framework of this paper, replacing at each simplex in the Regge path integral the exponential of the action by its spinfoam-inspired equivalent, and showing that the contribution from the non-Regge terms is suppressed in the semiclassical limit\footnote{This can be seen as a continuation and generalization of the work done in \cite{Barrett:1993db} for the Ponzano-Regge model in 3d. See also \cite{Livine:2002rh,Oriti:2004mu,Oriti:2005jr,Oriti:2006wq} for a different perspective on the ``cosine problem'' that traces it to the need to implement causality in the model.}.

Let us first state some expectations for the semiclassical behaviour of the vertex amplitude of a spin foam model (for general reviews of the spin foam framework, see \cite{Baez:1997zt,Oriti:2001qu,Perez:2003vx}). A spin foam vertex amplitude $A_v(j_f,i_e)$ for a four-dimensional model depends on the spins $j_f$ and the intertwiners $i_e$ that colour the ten faces and five edges surrounding the vertex; these faces and edges are dual to the triangles and tetrahedra surrounding the simplex dual to $v$ in a simplicial triangulation of the manifold. When the spins $j_f$ are sent to infinity and the intertwiners $i_e$ are chosen to be a specific function of the spins, the vertex amplitude is conjectured to have the following asymptotic behaviour:
\begin{equation}\label{PcosD}
 A_v(j_f,\,i_e)\sim P_v \cos\left[\left(\gamma\sum_{j_f} j_f \theta_f\right)+\frac{\pi}{4} \right] +D_v\;,\quad\quad j_f \to\infty\;.
\end{equation}
In this expression, $P_v$ scales as $j_0^{-m}$ when all spins are rescaled by a factor $j_0$. The coefficients $\theta_f$ do not scale with $j_0$ and each has the appropriate value for the internal dihedral angle at the triangle dual to face $f$ in the simplex dual to $v$. Newton's constant enters in the argument of the cosine through the identification of the spins $j_f$ with the areas of the triangles by the LQG formula $A_a=\hbar\kappa\gamma\, j_f$ for the large spin asymptotics of the spectrum of the area \cite{Rovelli:1994ge,Ashtekar:1996eg}. Note that the argument of the cosine is therefore the Regge action for the simplex dual to $v$ (divided by $\hbar$)\footnote{Recall that, in order to obtain a deficit angle, we have to subtract to $2\pi$ the sum of the dihedral angles at a triangle. As a result, to recover the Regge action from a sum over simplices, an extra term $2\pi A_a$ for each triangle is needed.  This term is expected to arise from the face amplitude of the spin foam model.}.  The term $D_v$, finally, does not have any ``good'' interpretation as a classical geometry. It scales as $j_0^{-d}$ with $d<m$, and is therefore dominant over the cosine term, but it contrasts with it in being slowly varying (not oscillating). The presence of a cosine, of the $\pi/4$ and of the dominant non-oscillating $D_v$ term are standard features of asymptotic expansions. The form (\ref{PcosD}) of the asymptotic vertex amplitude is known to be true for the Barrett-Crane (BC) model, with $m=9/2$ and $d=2$  \cite{Barrett:1998gs,Baez:2002rx,Barrett:2002ur,Freidel:2002mj}. This is the main motivation for us considering it, under the conjecture (made reasonable by recent developments discussed in the next section) that the more recent spin foam models \cite{Engle:2007uq,Livine:2007vk,Engle:2007qf,Freidel:2007py,Engle:2007wy} lead to identical or similar behaviour. The only addition we have made to the BC result is the Immirzi parameter $\gamma$ to scale the areas correctly in accordance with LQG; since the new models, unlike BC, include $\gamma$ and reproduce the LQG area spectrum \cite{Engle:2007wy}, this seems a reasonable assumption.

We want to focus our investigation on the consequences of having such an expression at each simplex of the triangulation. We therefore examine the following definition of observables:
\begin{align}\label{obscos}
 \langle \hat{\mathcal{O}}\rangle &= \frac{1}{Z_q}\int \prod_{i,n} \mathrm{d}l_n\,\mathrm{d}L_i\,
\tilde{\mu}(l_n, L_i)\,\,\prod_\sigma \left[ P_\sigma(l_n^\sigma,L_i^\sigma)\cos\left(\frac{1}{\hbar}S_\sigma(l_n^\sigma,L_i^\sigma)+\frac{\pi}{4}\right)+D_\sigma(l_n^\sigma,L_i^\sigma)\right]  \nonumber\\
& \hat{\mathcal{O}}\, \,\exp\left[-\frac{1}{2 \hbar \kappa}\,\alpha_{ij}\,(L_i-L_i^0)\, (L_j-L_j^0)\right] \exp\left[-\frac{i}{\hbar}\,K_i^0\,(L_i-L_i^0)\right] \;.
\end{align}
This is identical to the Regge calculus expression (\ref{obs22}), with the exponential of the action replaced, at each simplex $\sigma$, by the conjectured limit of spin foam models. The dependence of $S_\sigma$, $P_\sigma$ and $D_\sigma$ has been translated from the spins to the lengths via the areas; $(l_n^\sigma,L_i^\sigma)$ are the internal and (if any) boundary lengths belonging to simplex $\sigma$, and we write $\tilde{\mu}$ for the measure because it may be different from the pure Regge measure $\mu$. There are several hypothesis and assumptions needed for using (\ref{obscos}) as the correct semiclassical regime of spin foam observables; the plausibility of them will be discussed and defended in the next section, but for the moment (\ref{obscos}) can be taken as an ansatz for focusing on the effects of the ``$P \cos\, S+ D$'' structure and studying whether they hamper the correct semiclassical limit.

The product over simplices can be expanded using Euler's formula leading to
\begin{align}\label{sumeps}
\langle \hat{\mathcal{O}} \rangle &= \frac{1}{Z_q}\int \prod_{m,n} \mathrm{d}l_n\,\mathrm{d}L_m\,
 \tilde{\mu}(l_n, L_i)\,\,\sum_\epsilon \left[\left(\prod_\sigma P_{\epsilon}^{\sigma}(l_n^\sigma,L_i^\sigma) \right)  \exp\left( \frac{i}{\hbar} \sum_\sigma S_\epsilon^\sigma(l_n^\sigma,L_i^\sigma)\right) \right] \nonumber\\
& \hat{\mathcal{O}}\,\,\exp\left[-\frac{1}{2 \hbar \kappa}\,\alpha_{ij}\,(L_i-L_i^0)\, (L_j-L_j^0)\right] \exp\left[-\frac{i}{\hbar}\,K_i^0\,(L_i-L_i^0)\right] \; .
\end{align}
Here $\epsilon$ runs over the values $\{+1, -1, 0\}$ for each simplex of the triangulation, and we have defined:
\begin{subequations}
\begin{align}
S_{+1}^\sigma &= S_\sigma &P_{+1}^\sigma =P_\sigma\, \mathrm{e}^{\frac{i\pi}{4}} \\
S_{-1}^\sigma &= -S_\sigma &P_{-1}^\sigma =P_\sigma\, \mathrm{e}^{-\frac{i\pi}{4}}   \\
S_{0}^\sigma &= 0 &P_{0}^\sigma =D_\sigma
\end{align}
\end{subequations}

Before applying the stationary phase approximation for small $\hbar$ to each of these terms, we need to make explicit the hidden dependence on $\hbar$ in the scaling\footnote{In principle $\tilde{\mu}$ could also scale with $\hbar$, but this scaling is irrelevant since it is an overall factor that cancels when dividing by $Z_q$.} of $P_\epsilon^\sigma$. Since $\hbar$ appears when we translate spins to areas with the LQG equivalence $A=\hbar\kappa\gamma\, j$, the scaling with $\hbar$ of $P_\epsilon^\sigma$ is inverse to its scaling with $j_0$ in the large spin limit:
\begin{subequations}
\begin{align}
&P_\pm^\sigma \sim \hbar^{-m}\tilde{P}_\pm^\sigma \\
&P_{0}^\sigma \sim \hbar^{-d}\tilde{P}_{0}^\sigma\,,\quad\quad\quad\quad\quad 0<d<m\;,
\end{align}
\end{subequations}
where $\tilde{P}_\epsilon^\sigma$ is independent of $\hbar$ in the limit under consideration.

Let us examine the numerator of (\ref{sumeps}). It is a sum of $3^k$ terms -- $k$ being the total number of simplices in the triangulation -- where each term corresponds to an assignment $\epsilon=\{\epsilon_\sigma\}$ to each simplex and has the general form
\begin{equation}\label{terms}
\hbar^{-q} \int \prod_{m,n} \mathrm{d}l_n\,\mathrm{d}L_m\,
 f_\epsilon(l_n,L_i)\, \mathcal{O}(L_m)\, \exp\left[ \frac{i}{\hbar}Q_\epsilon(l_n,L_i)\right]\;.
\end{equation}
The power $q$ equals the sum of the powers $m$ or $d$ coming from each simplex, according to the assignment of signs for the term under consideration, whereas $f_\epsilon(l_n,L_m)$ is given by $\tilde{\mu}$ times a product over simplices of $P_{\epsilon}^{\sigma}$ and the phase function is
\begin{equation}
Q_\epsilon(l_n,L_i)= S_\epsilon(l_n,L_i) + \frac{i}{2\kappa}\,\alpha_{ij}\left(L_i-L_i^0\right)\left(L_{j}-L_{j}^0\right)-\,K_i^0\,\left(L_i-L_i^0\right)\;.
\end{equation}
The ``action'' $S_\epsilon(l_n,L_i)$ has been defined as the sum over simplices of $S_\epsilon^\sigma$. For each term of the form (\ref{terms}), we can now apply the stationary phase expansion for small $\hbar$ (which as we know is physically equivalent to large boundary lengths compared to $l_P$). Just as in the analogous calculation in subsection \ref{sec:2.2}, the imaginary part of $Q_\epsilon$ is minimized when the boundary lengths attain the classical values peaked on by the boundary state, $L_i=L_i^0$. This selects a region $D_0$ in the configuration space $\{l_n,L_i\}$, and the question is now whether the real part of $Q$ has an extremum in this region. The location of such an extremum would be given by the solution to the equations
\begin{subequations}
\begin{align}
\frac{\partial}{\partial l_n}S_\epsilon(l_n,L_m^0)& =\;  0 \quad. \label{extremum2SF}\\
\left.\frac{\partial}{\partial L_i}S_\epsilon(l_n,L_i)\right|_{L_i=L_i^0}&=\;K_i^0  \;,\label{extremum3SF}
\end{align}
\end{subequations}
These equations are analogous to (\ref{extremum2},\ref{extremum3}).  Equation (\ref{extremum2SF}) is a Dirichlet boundary value problem for the action $S_\epsilon(l_n,L_i^0)$. On the other hand, equation (\ref{extremum3SF}) is a Neumann boundary condition for the same action.  For the particular case where $\epsilon=+1$ at each simplex, the set of equations (\ref{extremum2SF}-\ref{extremum3SF}) has a consistent solution -- the classical Regge solution determined by the boundary state -- because the phase factor $K_i^0$ is hand-picked to guarantee this consistency;
 but for a general term, the set of values $\{l_n^0\}$ that satisfy the first equation (solution to the equations of motion for an action with ``wrong'' signs at different simplices) is not going to satisfy the second one as well. Hence for a general term in the sum we fall into case (i) or (ii) of the three possibilities enumerated after (\ref{eq:Hessian}), and the contribution from the term is suppressed faster than any finite power of $\hbar$. Notice that this happens regardless of the power $q$ that comes with $\hbar$ in front of the term, therefore the dominance of the degenerate configurations in the limit of the vertex amplitude is now irrelevant; the semiclassical boundary state ensures that the contribution from the degenerate terms, as well as the sign reversed ones, vanishes from the semiclassical evaluation of boundary observables.

the argument above makes a non-trivial assumption about the measure $f_\epsilon(l_n,L_i)$ arising from the spin foam model: we have assumed that the measure vanishes smoothly on the boundary of the domain of integration. If this does not happen (i.e. if the measure gives non-zero weight to length values that saturate triangular inequalities) then the asymptotic expansion features a boundary term which is subdominant but not exponentially suppressed. As a result, restricting attention to terms with $\epsilon=\pm1$ only at each simplex, a milder result holds: the ``cosines'' give contributions only at next-to-leading order. On the other hand, without further assumptions, the contributions coming from terms with $D_\sigma$ factors could dominate and destroy the semiclassical behaviour.

In summary, within our assumptions, the only non-suppressed term from the numerator is the one with an action exactly equal to the Regge action for the whole triangulation, with no contribution from sign-reversed or ``degenerate'' terms. The same thing happens in the denominator $Z_q$, and thus the evaluation of observables matches that of conventional Regge calculus. The agreement is exact for the lowest order in the one-point and connected two-point functions, which are independent of the measure; it becomes exact at all higher orders as well if we posit the identification of measures
\begin{equation}
 f_+(l_n,L_i):=\tilde{\mu}(l_n, L_i)\prod_\sigma P_{+1}^{\sigma}(l_n^\sigma,L_i^\sigma)=\mu(l_n,L_i)\;.
\end{equation}
This result is a strong indication that spin foam models might be able to recover the correct semiclassical limit, since the presence of the ``$P \cos\,S + D$'' factor at each simplex instead of the exponential of the Regge action does not affect the semiclassical evaluation of observables. 

\section{Discussion}\label{sec:4}

The main aim of this paper has been to make clearer and cleaner the application of the boundary state formalism for obtaining the semiclassical regime of nonperturbative quantum-gravitational theories. We have focused on Regge calculus because, as a version of general relativity with a finite number of degrees of freedom and a better-defined and relatively well-understood quantum theory, it provides an ideal testing ground for the formalism. 
Moreover, Regge calculus is intimately related to spin foam models both in the way they are defined (from a simplicial discretization of BF theory with constraints) and in their conjectured semiclassical limit (through the relation of the vertex amplitude to the cosine of the Regge simplicial action). 
The main results of the paper are simply summarized as: 
\begin{itemize}
	\item the semiclassical boundary state for Regge calculus successfully selects the semiclassical regime for the whole path integral, thereby bridging the gap between nonperturbative and perturbative theory for a general triangulation;
	\item when at each simplex we replace the exponential of the Regge action by the cosine which is conjectured to come from the relevant limit of spin foam models, the results are unchanged because the non-Regge terms are suppressed from the semiclassical regime.
\end{itemize}
The result we obtain for the connected two-point function of boundary edge lengths, a key object related to the simplicial graviton propagator, is
\begin{equation}\label{hamilton-conc}
\langle\, L_i\, L_j \,\rangle_q - \langle\, L_i\, \rangle_q \,\langle\, L_j\, \rangle_q  = \hbar \, \Big(\frac{\p^2 S^H}{\p L_i \p L_j}(L_i^0)+\frac{i}{\kappa} \alpha_{ij}\Big)^{-1}\;,
\end{equation}
with $S^H$ being the Hamilton function of the configuration and $\alpha$ the correlation matrix of the boundary state.

The approach we have followed mirrors the ``graviton propagator'' calculations of \cite{Rovelli:2005yj,Bianchi:2006uf}. The key technique in our analysis is to do the integral on internal and on boundary variables at once. The integral contains a Gaussian boundary state peaked on a classical configuration and we take the limit in which the relative dispersion of intrinsic and extrinsic geometry in the Gaussian vanish. The parameter encoding this ``semiclassicality'' is the ratio of the typical length scale of the classical solution peaked upon to the Planck scale. The stationary phase approximation, first done formally as an $\hbar\to 0$ limit, is revealed to be in fact an expansion in inverse powers of this large parameter.

Having identified the parameter encoding semiclassicality allows to develop a \emph{semiclassical perturbation theory}. The results presented in this paper are restricted to one- and two-point functions at the leading order, but the formalism allows us to compute both $n$-point functions and quantum corrections to the classical behaviour. The quantum corrections are organized in a loop expansion akin to the one used in ordinary quantum field theory with a background. The difference here is that the ``background'' geometry emerges from the nonperturbative theory only in a specific regime selected by the boundary state. \\

Given the fact that the semiclassical boundary state plays such a key role in this approach, one can naturally ask whether it is possible to derive it from first principles and especially derive its free parameters $\alpha_{ij}$. A comparison of the conventional path integral over all lengths (\ref{Z-def}) and our version that integrates only the interior and boundary of a region (\ref{Z-int}) suggests that the boundary quantum state can be defined as an integral over exterior variables:
\begin{equation}\label{def-Psi}
  \Psi(L_i)=\int \prod_k \mathrm{d}l_k^{\mathrm{ext}}\,\mu(l_k^{\mathrm{ext}},L_i)\,\exp\left[  \frac{i}{\hbar} S^{\mathrm{ext}}(l_k^{\mathrm{ext}},L_i)\right] \;.
\end{equation}
A possibility of this kind is discussed for instance in \cite{Dittrich:2007wm}. From this point of view, a choice of $\Psi$ is equivalent to a choice of asymptotic boundary conditions in the full path integral. However a difficulty can be easily identified in this attempt: the $\alpha_{ij}$ coming from the external integration is purely imaginary, therefore it cannot have a positive definite real part. This fact suggests that, even when considering the external integral, an \emph{asymptotic} boundary state is needed\footnote{As a side remark we would like to recall that, in the standard approach to perturbative quantum field theory where states are associated to spatial sections, the vacuum state $\Psi_0$ at a given time $t$ is also a Gaussian functional with a real positive definite $\alpha$. It can be obtained from the partition function $Z_0$ by integrating out from time $t$ to $-\infty$; however, unless a Wick rotation is implemented, this would give an imaginary $\alpha$. This shows that even in ordinary quantum field theory the partition function $Z_0$ contains a boundary vacuum state with real $\alpha$ (defined asymptotically at $t=\pm\infty$). This issue is discussed in chapter 9 of Weinberg's \cite{Weinberg:1995mt} and is the origin of the $+i \epsilon$ prescription.}. A standard trick to obtain the vacuum state in conventional quantum mechanics and field theory is to perform a Wick rotation. However the implementation of a meaningful analogue of this procedure (and even the existence of one) in a general-boundary setting is still unclear\footnote{In \cite{Oeckl:2003vu,Oeckl:2005bv,Oeckl:2005bw,Colosi:2007bj,Colosi:2008fv} the boundary vacuum state is found starting with a Gaussian ansatz and then imposing a set of consistency conditions.}.

A more promising route might be to select the boundary state representing the semiclassical regime in an a posteriori way, by \textit{requiring} that the results match those of conventional perturbative Regge theory (in which only small fluctuations around a given classical solution for the whole manifold are quantized and summed over; see \cite{Rocek:1982fr,Rocek:1982tj,Hamber:2004ew}). Clearly this vindicates the general Gaussian expression to the lowest order, and the free parameters $\alpha_{ij}$ in this parametrization (one for each pair of boundary edges $L_i,L_j$) can be established a posteriori by matching the correlations between each pair of boundary edges with the conventional perturbative expression for it. 
A nontrivial consistency check of this matching can be found if we consider deformations of the boundary. Given a Gaussian boundary state for a given boundary including edges $I$ and $J$, if we choose a different boundary which also includes these two edges we can find (by integrating out from (\ref{def-Psi}) the edges external to one boundary and not to the other one) how the new coefficient $\alpha_{IJ}'$ in the new boundary state is related to the old one $\alpha_{IJ}$. On the other hand, the requirement that the correlation between two edges is invariant under the change of boundary (which must be true to match conventional perturbative theory, which knows nothing about boundaries) gives a separate constraint on the variation of the $\alpha$ matrix, and the consistency of the two constraints is a nontrivial check on the validity of the approach and its ability to recover conventional perturbative theory correctly. We hope to discuss these matters more fully in a future paper.\\
 
Moving on to the results of Section \ref{sec:2.2}, we believe that they may help clarify an important aspect of the ``gravitons from spinfoams'' calculation: namely Rovelli's suggestion that few spinfoam vertices can provide a good approximation for low energy processes \cite{Rovelli:2008la}. We find that, in the flat space case for quantum Regge calculus, connected two-point correlation functions are, at leading order, independent of the triangulation. This means that, given a triangulation of the boundary, a minimal triangulation of the bulk (one with no internal vertices, so that  only a number of bulk diagonals is present) is enough to obtain the exact result. Triangulation independence is expected to be lost at next-to-leading order; as a result we would have that quantum corrections to simplicial flat space correspond to unfreezing the degrees of freedom associated to the internal vertices of the triangulation.\\

Quantum Regge calculus provides an ideal arena for testing the ideas introduced  in \cite{Rovelli:2005yj,Bianchi:2006uf}. Moreover, the results presented in this paper (and in particular in section \ref{sec:3}) may turn out to be directly relevant for the analysis of the semiclassical regime of the new spin foam models \cite{Engle:2007uq,Livine:2007vk,Engle:2007qf,Freidel:2007py,Engle:2007wy} beyond the single vertex level, if a number of gaps are filled. 
The first one is obviously whether their large spin limit is in fact related to the cosine Regge expression for each simplex as postulated in expression (\ref{PcosD}). There are strong indications that this is likely to be true. On one side, recent numerical and analytical work on the so-called ``flipped'' vertex \cite{Magliaro:2007nc,abmp:2008n} indicates that it has a well-behaved semiclassical regime with a geometrical interpretation; on the other, a stationary phase analysis of the Freidel-Krasnov (FK)  model \cite{Freidel:2007py} after writing it in path integral form \cite{Conrady:2008ea,New2} finds not only that the semiclassical limit of the vertex is the cosine of the Regge action, but also that the mismatch between area and length variables is side-stepped since area configurations that do not correspond to simplicial geometries drop out from the semiclassical limit. This last development is especially encouraging since the variable mismatch \cite{Dittrich:2008va} and the presence of discontinuous boundary geometries \cite{Bianchi:2008es,Dittrich:2008ar} is another general problem the spin foam formalism seems to face when trying to obtain a semiclassical limit with a clear geometrical interpretation. 

The recent formulation of the FK model as a path integral \cite{Conrady:2008ea} also opens the door for the possibility of doing the stationary phase analysis with a boundary state in a purely spin foam context. Hopefully this would vindicate the implicit assumption in our ``hybrid'' expression (\ref{PcosD}) that the semiclassical ``$\textrm{cos}\, S + D$'' expression for the vertex amplitude can be taken at each internal vertex of the two-complex, even if the state sum is nonperturbative and the semiclassical regime is enforced with a boundary state that peaks explicitly only the boundary variables. It seems likely that this could be done by adding a suitable boundary state to the formal stationary phase analysis of the FK model mentioned in the previous paragraph.\\

We believe that the results of section \ref{sec:3} about the dropping out of the all the non-Regge terms coming from the conjectured limit of the spin foam vertex amplitude is very encouraging news for the spin foam research program. The ``cosine problem'' has been a potential obstacle for the semiclassical limit of spin foams from the very beginning of this research program, as can be seen already in Ponzano and Regge's 1968 paper \cite{PonzanoRegge:1968}, where they say referring to the ``non-Regge'' terms arising from the large spin limit in 3d gravity: ``\emph{The other terms, other than the positive frequency part, are related to different orientations of the tetrahedra $T_j$ and have a similar interpretation, although their precise meaning is still unclear. It is plausible that in the transition to a smooth manifold $\mathcal{M}$ they will give no essential contribution to the final result}''. Our results show that if the large spin limit of a spin foam state sum is a Regge-like path integral with the cosine factor at each simplex, Ponzano and Regge's hope is realized and the non-positive frequency terms give no contribution to the final result, once the desired semiclassical regime is selected with an appropriate boundary state.

\section*{Acknowledgments}
\hspace{1.5em} We thank Carlo Rovelli for valuable suggestions and for comments on an early draft of this work, and Simone Speziale and Bianca Dittrich for helpful comments on a more recent draft. We also wish to thank Ruth Williams, Laurent Freidel, John Barrett, Jorma Louko, Tamer Tlas, Florian Conrady, Daniele Oriti, Leonardo Modesto and Jonathan McDonald for helpful discussions. A.S. was supported by an ESF Exchange Visit Travel Grant. E.B. was supported by a Della Riccia Fellowship.


\begin{thebibliography}{10}

\bibitem{Rovelli:2004tv}
C.~Rovelli, ``{Quantum gravity}''. Cambridge, UK: Univ. Pr. (2004) 455 p.

\bibitem{Baez:1997zt}
J.~C. Baez, ``{Spin foam models}'' {\em Class. Quant. Grav.} {\bf 15} (1998)
  1827--1858,
\href{http://arXiv.org/abs/gr-qc/9709052}{{\tt gr-qc/9709052}}.

\bibitem{Oriti:2001qu}
D.~Oriti, ``{Spacetime geometry from algebra: Spin foam models for non-
  perturbative quantum gravity}'' {\em Rept. Prog. Phys.} {\bf 64} (2001)
  1489--1544,
\href{http://arXiv.org/abs/gr-qc/0106091}{{\tt gr-qc/0106091}}.

\bibitem{Perez:2003vx}
A.~Perez, ``{Spin foam models for quantum gravity}'' {\em Class. Quant. Grav.}
  {\bf 20} (2003) R43,
\href{http://arXiv.org/abs/gr-qc/0301113}{{\tt gr-qc/0301113}}.

\bibitem{Donoghue:1994dn}
J.~F. Donoghue, ``{General relativity as an effective field theory: The leading
  quantum corrections}'' {\em Phys. Rev.} {\bf D50} (1994) 3874--3888,
\href{http://arXiv.org/abs/gr-qc/9405057}{{\tt gr-qc/9405057}}.

\bibitem{Burgess:2003jk}
C.~P. Burgess, ``{Quantum gravity in everyday life: General relativity as an
  effective field theory}'' {\em Living Rev. Rel.} {\bf 7} (2004) 5,
\href{http://arXiv.org/abs/gr-qc/0311082}{{\tt gr-qc/0311082}}.

\bibitem{Ashtekar:2006yw}
A.~Ashtekar, ``{Physics from geometry}'' {\em Nature Phys.} {\bf 2} (2006)
725--726.

\bibitem{Rovelli:2005yj}
C.~Rovelli, ``{Graviton propagator from background-independent quantum
  gravity}'' {\em Phys. Rev. Lett.} {\bf 97} (2006) 151301,
\href{http://arXiv.org/abs/gr-qc/0508124}{{\tt gr-qc/0508124}}.

\bibitem{Bianchi:2006uf}
E.~Bianchi, L.~Modesto, C.~Rovelli, and S.~Speziale, ``{Graviton propagator in
  loop quantum gravity}'' {\em Class. Quant. Grav.} {\bf 23} (2006)
  6989--7028,
\href{http://arXiv.org/abs/gr-qc/0604044}{{\tt gr-qc/0604044}}.

\bibitem{Thiemann:2007zz}
T.~Thiemann, ``{Modern canonical quantum general relativity}''. Cambridge, UK:
  Cambridge Univ. Pr. (2007) 819 p.

\bibitem{Ashtekar:2004eh}
A.~Ashtekar and J.~Lewandowski, ``{Background independent quantum gravity: A
  status report}'' {\em Class. Quant. Grav.} {\bf 21} (2004) R53,
\href{http://arXiv.org/abs/gr-qc/0404018}{{\tt gr-qc/0404018}}.

\bibitem{Barrett:1998gs}
J.~W. Barrett and R.~M. Williams, ``The asymptotics of an amplitude for the
  4-simplex'' {\em Adv. Theor. Math. Phys.} {\bf 3} (1999) 209--215,
\href{http://arXiv.org/abs/gr-qc/9809032}{{\tt gr-qc/9809032}}.

\bibitem{Oeckl:2003vu}
R.~Oeckl, ``{A `general boundary' formulation for quantum mechanics and quantum
  gravity}'' {\em Phys. Lett.} {\bf B575} (2003) 318--324,
\href{http://arXiv.org/abs/hep-th/0306025}{{\tt hep-th/0306025}}.

\bibitem{Oeckl:2005bv}
R.~Oeckl, ``{General boundary quantum field theory: Foundations and probability
  interpretation}'' {\em Adv. Theor. Math. Phys.} {\bf 12} (2008) 319--352,
\href{http://arXiv.org/abs/hep-th/0509122}{{\tt hep-th/0509122}}.

\bibitem{Oeckl:2005bw}
R.~Oeckl, ``{General boundary quantum field theory: Timelike hypersurfaces in
  Klein-Gordon theory}'' {\em Phys. Rev.} {\bf D73} (2006) 065017,
\href{http://arXiv.org/abs/hep-th/0509123}{{\tt hep-th/0509123}}.

\bibitem{Colosi:2007bj}
D.~Colosi and R.~Oeckl, ``{S-matrix at spatial infinity}'' {\em Phys. Lett.}
  {\bf B665} (2008) 310--313,
\href{http://arXiv.org/abs/0710.5203}{{\tt 0710.5203}}.

\bibitem{Colosi:2008fv}
D.~Colosi and R.~Oeckl, ``{Spatially asymptotic S-matrix from general boundary
  formulation}''
\href{http://arXiv.org/abs/0802.2274}{{\tt 0802.2274}}.

\bibitem{Rovelli:1994ge}
C.~Rovelli and L.~Smolin, ``{Discreteness of area and volume in quantum
  gravity}'' {\em Nucl. Phys.} {\bf B442} (1995) 593--622,
\href{http://arXiv.org/abs/gr-qc/9411005}{{\tt gr-qc/9411005}}.

\bibitem{Ashtekar:1996eg}
A.~Ashtekar and J.~Lewandowski, ``{Quantum theory of geometry. I: Area
  operators}'' {\em Class. Quant. Grav.} {\bf 14} (1997) A55--A82,
\href{http://arXiv.org/abs/gr-qc/9602046}{{\tt gr-qc/9602046}}.

\bibitem{Ashtekar:1997fb}
A.~Ashtekar and J.~Lewandowski, ``{Quantum theory of geometry. II: Volume
  operators}'' {\em Adv. Theor. Math. Phys.} {\bf 1} (1998) 388--429,
\href{http://arXiv.org/abs/gr-qc/9711031}{{\tt gr-qc/9711031}}.

\bibitem{Major:1999mc}
S.~A. Major, ``{Operators for quantized directions}'' {\em Class. Quant.
  Grav.} {\bf 16} (1999) 3859--3877,
\href{http://arXiv.org/abs/gr-qc/9905019}{{\tt gr-qc/9905019}}.

\bibitem{Thiemann:1996at}
T.~Thiemann, ``{A length operator for canonical quantum gravity}'' {\em J.
  Math. Phys.} {\bf 39} (1998) 3372--3392,
\href{http://arXiv.org/abs/gr-qc/9606092}{{\tt gr-qc/9606092}}.

\bibitem{Bianchi:2008es}
E.~Bianchi, ``{The length operator in Loop Quantum Gravity}''
  \href{http://arXiv.org/abs/0806.4710}{{\tt 0806.4710}}. \emph{Nucl. Phys.} \textbf{B} in press. \href{http://dx.doi.org/10.1016/j.nuclphysb.2008.08.013}{{\tt dx.doi.org/10.1016/j.nuclphysb.2008.08.013}}

\bibitem{Barrett:1997gw}
J.~W. Barrett and L.~Crane, ``Relativistic spin networks and quantum gravity''
  {\em J. Math. Phys.} {\bf 39} (1998) 3296--3302,
\href{http://arXiv.org/abs/gr-qc/9709028}{{\tt gr-qc/9709028}}.

\bibitem{Livine:2006it}
E.~R. Livine and S.~Speziale, ``{Group integral techniques for the spinfoam
  graviton propagator}'' {\em JHEP} {\bf 11} (2006) 092,
\href{http://arXiv.org/abs/gr-qc/0608131}{{\tt gr-qc/0608131}}.

\bibitem{Christensen:2007rv}
J.~D. Christensen, E.~R. Livine, and S.~Speziale, ``{Numerical evidence of
  regularized correlations in spin foam gravity}''
\href{http://arXiv.org/abs/0710.0617}{{\tt 0710.0617}}.

\bibitem{Bianchi:2007vf}
E.~Bianchi and L.~Modesto, ``{The perturbative Regge-calculus regime of Loop
  Quantum Gravity}'' {\em Nucl. Phys.} {\bf B796} (2008) 581--621,
\href{http://arXiv.org/abs/0709.2051}{{\tt 0709.2051}}.

\bibitem{Alesci:2007tx}
E.~Alesci and C.~Rovelli, ``{The complete LQG propagator: I. Difficulties with
  the Barrett-Crane vertex}'' {\em Phys. Rev.} {\bf D76} (2007) 104012,
\href{http://arXiv.org/abs/0708.0883}{{\tt 0708.0883}}.

\bibitem{Alesci:2007tg}
E.~Alesci and C.~Rovelli, ``{The complete LQG propagator: II. Asymptotic
  behavior of the vertex}'' {\em Phys. Rev.} {\bf D77} (2008) 044024,
\href{http://arXiv.org/abs/0711.1284}{{\tt 0711.1284}}.

\bibitem{Engle:2007uq}
J.~Engle, R.~Pereira, and C.~Rovelli, ``{The loop-quantum-gravity
  vertex-amplitude}'' {\em Phys. Rev. Lett.} {\bf 99} (2007) 161301,
\href{http://arXiv.org/abs/0705.2388}{{\tt 0705.2388}}.

\bibitem{Livine:2007vk}
E.~R. Livine and S.~Speziale, ``{A new spinfoam vertex for quantum gravity}''
  {\em Phys. Rev.} {\bf D76} (2007) 084028,
\href{http://arXiv.org/abs/0705.0674}{{\tt 0705.0674}}.

\bibitem{Engle:2007qf}
J.~Engle, R.~Pereira, and C.~Rovelli, ``{Flipped spinfoam vertex and loop
  gravity}'' {\em Nucl. Phys.} {\bf B798} (2008) 251--290,
\href{http://arXiv.org/abs/0708.1236}{{\tt 0708.1236}}.

\bibitem{Freidel:2007py}
L.~Freidel and K.~Krasnov, ``{A New Spin Foam Model for 4d Gravity}'' {\em
  Class. Quant. Grav.} {\bf 25} (2008) 125018,
\href{http://arXiv.org/abs/0708.1595}{{\tt 0708.1595}}.

\bibitem{Engle:2007wy}
J.~Engle, E.~Livine, R.~Pereira, and C.~Rovelli, ``{LQG vertex with finite
  Immirzi parameter}'' {\em Nucl. Phys.} {\bf B799} (2008) 136--149,
\href{http://arXiv.org/abs/0711.0146}{{\tt 0711.0146}}.

\bibitem{PonzanoRegge:1968}
G.~Ponzano and T.~Regge, ``Semiclassical limit of Racah coeffecients''. in
  Spectroscopic and Group Theoretical Methods in Physics, edited by F. Block
  (North Holland, Amsterdam, 1968).

\bibitem{Regge:1961px}
T.~Regge, ``General Relativity without coordinates'' {\em Nuovo Cim.} {\bf 19}
  (1961)
558--571.

\bibitem{Loll:1998aj}
R.~Loll, ``Discrete approaches to quantum gravity in four dimensions'' {\em
  Living Rev. Rel.} {\bf 1} (1998) 13,
\href{http://arXiv.org/abs/gr-qc/9805049}{{\tt gr-qc/9805049}}.

\bibitem{Regge:2000wu}
T.~Regge and R.~M. Williams, ``Discrete structures in gravity'' {\em J. Math.
  Phys.} {\bf 41} (2000) 3964--3984,
\href{http://arXiv.org/abs/gr-qc/0012035}{{\tt gr-qc/0012035}}.

\bibitem{Hamber:2007fk}
H.~W. Hamber, ``{Discrete and Continuum Quantum Gravity}''
\href{http://arXiv.org/abs/0704.2895}{{\tt 0704.2895}}.

\bibitem{Baez:2002rx}
J.~C. Baez, J.~D. Christensen, and G.~Egan, ``Asymptotics of 10j symbols''
  {\em Class. Quant. Grav.} {\bf 19} (2002) 6489,
\href{http://arXiv.org/abs/gr-qc/0208010}{{\tt gr-qc/0208010}}.

\bibitem{Barrett:2002ur}
J.~W. Barrett and C.~M. Steele, ``Asymptotics of relativistic spin networks''
  {\em Class. Quant. Grav.} {\bf 20} (2003) 1341--1362,
\href{http://arXiv.org/abs/gr-qc/0209023}{{\tt gr-qc/0209023}}.

\bibitem{Freidel:2002mj}
L.~Freidel and D.~Louapre, ``Asymptotics of 6j and 10j symbols'' {\em Class.
  Quant. Grav.} {\bf 20} (2003) 1267--1294,
\href{http://arXiv.org/abs/hep-th/0209134}{{\tt hep-th/0209134}}.

\bibitem{Hartle:1981cf}
J.~B. Hartle and R.~Sorkin, ``Boundary terms in the action for the Regge
  calculus'' {\em Gen. Rel. Grav.} {\bf 13} (1981)
541--549.

\bibitem{Friedberg:1984ma}
R.~Friedberg and T.~D. Lee, ``Derivation of Regge's action from Einstein's
  Theory of General Relativity'' {\em Nucl. Phys.} {\bf B242} (1984)
145.

\bibitem{MTW:1973}
C.~W. Misner, K.~S. Thorne, and J.~A. Wheeler, ``Gravitation''. (W. H.
  Freeman, 1973).

\bibitem{Wheeler:1964rgt}
J.~A. Wheeler, ``Geometrodynamics and the Issue of the Final State''. in
  Relativity, Groups and Topology, eds. B. DeWitt and C. DeWitt (New York,
  Gordon and Breach, 1964) 463-500.

\bibitem{Hamber:1997ut}
H.~W. Hamber and R.~M. Williams, ``On the measure in simplicial gravity'' {\em
  Phys. Rev.} {\bf D59} (1999) 064014,
\href{http://arXiv.org/abs/hep-th/9708019}{{\tt hep-th/9708019}}.

\bibitem{Jevicki:1985ta}
A.~Jevicki and M.~Ninomiya, ``Functional formulation of Regge gravity'' {\em
  Phys. Rev.} {\bf D33} (1986)
1634.

\bibitem{Menotti:1995ih}
P.~Menotti and P.~P. Peirano, ``Faddeev-Popov determinant in two-dimensional
  Regge gravity'' {\em Phys. Lett.} {\bf B353} (1995) 444--449,
\href{http://arXiv.org/abs/hep-th/9503181}{{\tt hep-th/9503181}}.

\bibitem{Menotti:1996tm}
P.~Menotti and P.~P. Peirano, ``Diffeomorphism invariant measure for finite
  dimensional geometries'' {\em Nucl. Phys.} {\bf B488} (1997) 719--734,
\href{http://arXiv.org/abs/hep-th/9607071}{{\tt hep-th/9607071}}.

\bibitem{Dittrich:2007wm}
B.~Dittrich, L.~Freidel, and S.~Speziale, ``{Linearized dynamics from the
  4-simplex Regge action}'' {\em Phys. Rev.} {\bf D76} (2007) 104020,
\href{http://arXiv.org/abs/0707.4513}{{\tt 0707.4513}}.

\bibitem{Speziale:2005ma}
S.~Speziale, ``{Towards the graviton from spinfoams: The 3d toy model}'' {\em
  JHEP} {\bf 05} (2006) 039,
\href{http://arXiv.org/abs/gr-qc/0512102}{{\tt gr-qc/0512102}}.

\bibitem{Livine:2006ab}
E.~R. Livine, S.~Speziale, and J.~L. Willis, ``{Towards the graviton from
  spinfoams: Higher order corrections in the 3d toy model}'' {\em Phys. Rev.}
  {\bf D75} (2007) 024038,
\href{http://arXiv.org/abs/gr-qc/0605123}{{\tt gr-qc/0605123}}.

\bibitem{Livine:2007mr}
E.~R. Livine and S.~Speziale, ``{Physical boundary state for the quantum
  tetrahedron}'' {\em Class. Quant. Grav.} {\bf 25} (2008) 085003,
\href{http://arXiv.org/abs/0711.2455}{{\tt 0711.2455}}.

\bibitem{Bonzom:2008xd}
V.~Bonzom, E.~R. Livine, M.~Smerlak, and S.~Speziale, ``{Towards the graviton
  from spinfoams: the complete perturbative expansion of the 3d toy model}''
\href{http://arXiv.org/abs/0802.3983}{{\tt 0802.3983}}.

\bibitem{Barrett:1994ks}
J.~W. Barrett {\em et al.}, ``{A Paralellizable implicit evolution scheme for
  Regge calculus}'' {\em Int. J. Theor. Phys.} {\bf 36} (1997) 815--840,
\href{http://arXiv.org/abs/gr-qc/9411008}{{\tt gr-qc/9411008}}.

\bibitem{Erdelyi:1954}
A.~Erdelyi, ``Asymptotic expansions''. (Dover Publications, New York, 1954).

\bibitem{vanderCorput:1954}
J.~van~der Corput, ``Asymptotic Expansions I. Fundamental theorems of
  Asymptotics''. (University of California, Berkeley, 1954).

\bibitem{Itzykson:1980rh}
C.~Itzykson and J.~B. Zuber, ``{Quantum Field Theory}''. New York, Usa:
  Mcgraw-hill (1980) (International Series In Pure and Applied Physics).

\bibitem{Rocek:1982fr}
M.~Rocek and R.~M. Williams, ``Quantum Regge calculus'' {\em Phys. Lett.} {\bf
  B104} (1981)
31.

\bibitem{Rocek:1982tj}
M.~Rocek and R.~M. Williams, ``The quantization of Regge calculus'' {\em Z.
  Phys.} {\bf C21} (1984)
371.

\bibitem{Hamber:2004ew}
H.~W. Hamber and R.~M. Williams, ``{Non-perturbative gravity and the spin of
  the lattice graviton}'' {\em Phys. Rev.} {\bf D70} (2004) 124007,
\href{http://arXiv.org/abs/hep-th/0407039}{{\tt hep-th/0407039}}.

\bibitem{Landau:1960}
L.~Landau and E.~Lifshits, ``Quantum Mechanics Non-Relativistic Theory''.
  (vol. 3, first ed., Pergamon, UK, 1958).

\bibitem{Dittrich:2008ar}
B.~Dittrich and J.~P. Ryan, ``{Phase space descriptions for simplicial 4d
  geometries}''
\href{http://arXiv.org/abs/0807.2806}{{\tt 0807.2806}}.

\bibitem{Barrett:1993db}
  J.~W.~Barrett and T.~J.~Foxon,
  ``{Semiclassical limits of simplicial quantum gravity}'',
  Class.\ Quant.\ Grav.\  {\bf 11} (1994) 543
  \href{http://arxiv.org/abs/gr-qc/9310016}{{\tt gr-qc/9310016}}.

\bibitem{Livine:2002rh}
E.~R. Livine and D.~Oriti, ``{Implementing causality in the spin foam quantum
  geometry}'' {\em Nucl. Phys.} {\bf B663} (2003) 231--279,
\href{http://arXiv.org/abs/gr-qc/0210064}{{\tt gr-qc/0210064}}.

\bibitem{Oriti:2004mu}
D.~Oriti, ``{The Feynman propagator for spin foam quantum gravity}'' {\em
  Phys. Rev. Lett.} {\bf 94} (2005) 111301,
\href{http://arXiv.org/abs/gr-qc/0410134}{{\tt gr-qc/0410134}}.

\bibitem{Oriti:2005jr}
D.~Oriti, ``{Generalised group field theories and quantum gravity transition
  amplitudes}'' {\em Phys. Rev.} {\bf D73} (2006) 061502,
\href{http://arXiv.org/abs/gr-qc/0512069}{{\tt gr-qc/0512069}}.

\bibitem{Oriti:2006wq}
D.~Oriti and T.~Tlas, ``{Causality and matter propagation in 3d spin foam
  quantum gravity}'' {\em Phys. Rev.} {\bf D74} (2006) 104021,
\href{http://arXiv.org/abs/gr-qc/0608116}{{\tt gr-qc/0608116}}.

\bibitem{Weinberg:1995mt}
S.~Weinberg, ``{The Quantum theory of fields. Vol. 1: Foundations}''.
  Cambridge, UK: Univ. Pr. (1995) 609 p.

\bibitem{Rovelli:2008la}
C.~Rovelli, talk given at $QG^2$ - Nottingham 2008. See also the discussion in
  section 3 of ref. \cite{Engle:2007wy}.

\bibitem{Magliaro:2007nc}
E.~Magliaro, C.~Perini, and C.~Rovelli, ``{Numerical indications on the
  semiclassical limit of the flipped vertex},'' {\em Class. Quant. Grav.} {\bf
  25} (2008) 095009,
\href{http://arXiv.org/abs/0710.5034}{{\tt 0710.5034}}.

\bibitem{abmp:2008n}
E.~Alesci, E.~Bianchi, E.~Magliaro, and C.~Perini, ``Intertwiner dynamics in
  the flipped vertex'', \href{http://arXiv.org/abs/0808.1971}{{\tt 0808.1971}}.


\bibitem{Conrady:2008ea}
F.~Conrady and L.~Freidel, ``{Path integral representation of spin foam models
  of 4d gravity}'', 
\href{http://arXiv.org/abs/0806.4640}{{\tt 0806.4640}}.

\bibitem{New2}
F.~Conrady and L.~Freidel,  ``{On the semiclassical limit of 4d spin foam models}'', \href{http://arxiv.org/abs/0809.2280}{{\tt 0809.2280}}.

\bibitem{Dittrich:2008va}
B.~Dittrich and S.~Speziale, ``{Area-angle variables for general relativity}'',
\href{http://arXiv.org/abs/0802.0864}{{\tt 0802.0864}}.

\end{thebibliography}


\providecommand{\href}[2]{#2}\begingroup\raggedright\endgroup


\end{document}